\documentclass[aps,prd,amsmath,amssymb,preprintnumbers,preprint,nofootinbib,a4paper,11pt]{revtex4}
\pdfoutput=1
\usepackage{graphicx}
\graphicspath{{./figures/}}
\usepackage{url}
\usepackage[bookmarks, pagebackref=true]{hyperref}
\usepackage{color}
	\definecolor{rossoCP3}{cmyk}{0,.88,.77,.40}
		\definecolor{graa}{rgb}{0.8,0.8,0.8}
		\definecolor{blaa}{rgb}{0.2,0.2,0.6}
		\hypersetup{
			colorlinks, 
			bookmarksopen, 
			bookmarksnumbered,
			citecolor=blaa, 		
			linkcolor=rossoCP3,	
			urlcolor=rossoCP3,			
			}
\usepackage{bm}
\usepackage{bbm}
\usepackage{pxfonts}
\usepackage[margin=5pt, font=small,labelfont=bf,justification=raggedright]{caption}
\usepackage{subfig}
\usepackage{youngtab}
\usepackage{slashed}

\newcommand{\beq}{\begin{eqnarray}}
\newcommand{\eeq}{\end{eqnarray}}

\newcommand{\ea}[1]{
\begin{align}
#1
\end{align}
}
\newcommand{\eas}[1]{
\begin{align*}
#1
\end{align*}
}

\newcommand{\bmp}{\noindent\begin{minipage}{16cm}}
\newcommand{\emp}{\end{minipage}\vskip 7mm} 

\newcommand{\id}{\mathrm{d}}

\newcommand{\Tr}{\text{Tr}}
\newcommand{\yf}{\tiny \yng(1)}

\begin{document}
\phantom{g}\vspace{2mm}
\title{ \LARGE  \color{rossoCP3} 
A Perturbative Realization of Miransky Scaling} 
\author{Oleg {\sc Antipin} 
}\email{antipin@cp3-origins.net}
\author{Stefano {\sc Di Chiara}
}\email{dichiara@cp3-origins.net}  
\author{Matin {\sc Mojaza}
}\email{mojaza@cp3-origins.net} 
\author{Esben {\sc M\o lgaard}
}\email{molgaard@cp3-origins.net} 
\author{Francesco {\sc Sannino}
}\email{sannino@cp3.dias.sdu.dk} 
\affiliation{
{ \color{rossoCP3}  \rm CP}$^{\color{rossoCP3} \bf 3}${\color{rossoCP3}\rm-Origins} \& the Danish Institute for Advanced Study {\color{rossoCP3} \rm DIAS},\ \\ 
University of Southern Denmark, Campusvej 55, DK-5230 Odense M, Denmark.
}
\begin{abstract}
Near conformal dynamics is employed in different extensions of the standard model of particle interactions as well as in cosmology. Many of its interesting properties are either conjectured or determined using model computations. We introduce a relevant four dimensional gauge theory template allowing us to investigate such dynamics perturbatively. The gauge theory we consider is quantum chromodynamics with the addition of a meson-like scalar degree of freedom as well as an adjoint Weyl fermion.  

At the two-loop level, and in the Veneziano limit, we firmly establish the existence of several fixed points of which one is all directions stable in the infrared. An interesting feature of the model is that this fixed point is lost, within the perturbatively trustable regime, by merging with another fixed point when varying the number of quark flavors. We show the emergence of the Miransky scaling and determine its properties. We are also able to determine the walking region of the theory which turns out to be, at large number of colors, about 12\%  of the conformal window. Furthermore, we determine highly relevant quantities for near conformal dynamics such as the anomalous dimension of the fermion masses.  
  \\[.1cm]
{\footnotesize  \it Preprint: CP$^3$-Origins-2012-15 \& DIAS-2012-16}
\end{abstract}
\maketitle

\newpage
\section{Introduction}
Determining the phase structure of gauge theories of fundamental interactions is crucial in order to be able to select relevant extensions of the standard model of particle interactions for particle physics, dark matter \cite{Sannino:2009za}, cosmology \cite{Channuie:2011rq,Bezrukov:2011mv} and quantum gravity \cite{Hooft:2010nc}. 

In particular  it is interesting to elucidate the perturbative and non-perturbative physics of non-Abelian gauge theories with and without fundamental scalars as function of the number of flavors, colors and matter representation. 

The relevant points to address are: 

\begin{itemize}
\item[i)]{Determine the regions of conformality of a given gauge theory as function of the couplings, number of flavors and matter representation.}

\item[ii)]{Determine whether near the conformal boundary between a conformally broken and conformally restored symmetry the theory {\it walks}  or {\it jumps}? Walking dynamics \cite{Holdom:1981rm,Holdom:1984sk,Yamawaki:1985zg,Appelquist:1986an} requires that the physical observables of the theory, in the broken phase, feel the presence of the conformal phase. This must happen at least in a region of the parameter space of the theory near the phase transition. Jumping dynamics \cite{Sannino:2012wy}  refers instead to the logical possibility that the theory in the broken phase displays no signs of a nearby, in the parameter space, conformal region. Jumping dynamics is associated to a first order conformal phase transition while walking dynamics to a second order or higher transition.} 

\item[iii)]{ Jumping and walking dynamics could be realized physically in different ways. For example walking arises naturally when an ultraviolet fixed point (UVFP) and an infrared one (IRFP) smoothly merge. The full beta function of the theory near this merger can be modeled by the product of two zeros. This mechanism also underlies the conjectured Miransky scaling  \cite{Miransky:1984ef,Miransky:1988gk,Miransky:1996pd} describing how the intrinsic scale of the theory vanishes as we approach the conformal boundary.  On the other hand the ratio of two zeros in the beta function leads naturally to jumping dynamics \cite{Sannino:2012wy}. The problem is therefore to establish which mechanism underlies either walking or jumping dynamics in a given gauge theory. }

\item[iv)]{Once the kind of the conformal transition has been established one needs to determine the physical properties of the theory. These are, for example, the anomalous dimensions of the fermion mass operator and the particle spectrum. }

\item[v)]{If the conformal transition is of walking type another question arises: What is the actual range of the parameter space of the theory supporting walking, e.g. the walking window as function of the number of flavors and colors? }

\end{itemize}

To help answer some of the points above it is, therefore, relevant to construct calculable examples such as the two dimensional toy model investigated in \cite{deForcrand:2012se}. Here, however, we are interested in investigating four dimensional gauge theories allowing us to address all, or part, of the questions above \cite{Grinstein:2011dq,Antipin:2011aa} in perturbation theory\footnote{Actually in \cite{Antipin:2011aa} we showed that one can also address some non-perturbative aspects.}. Since we are not concerned with the hierarchy problem, this is obviously true also for the toy model studied in \cite{deForcrand:2012se}, we consider theories featuring also fundamental scalars besides fermionic matter as done in  \cite{Grinstein:2011dq,Antipin:2011aa,Mojaza:2011rw}. 

We start by introducing the gauge theory template \cite{Antipin:2011aa} allowing us to investigate conformal dynamics and its breaking within perturbation theory. The gauge theory we consider is quantum chromodynamics  (QCD) with the addition of a meson-like scalar degree of freedom as well as an adjoint Weyl fermion. This model has a large number of global symmetries in common with ordinary  QCD  and allows us, at the two-loop level, and in the Veneziano limit, to establish the existence of several fixed points of which one is all directions stable in the infrared. 

A relevant feature of the model is that the infrared stable fixed point can be lost, within the perturbatively trustable regime, by either the merging with a non-trivial ultraviolet fixed point when varying the underlying number of quark species or via spontaneous condensation\footnote{We use the short-hand notation "condensation of the scalar field" to mean the spontaneous generation of its vacuum expectation value.} of a scalar field due to perturbative quantum corrections \cite{Coleman:1973jx}. Interestingly these two mechanisms associated to the loss of an infrared fixed point can be investigated in great detail within the same gauge theory. The merger phenomenon \cite{Gies:2005as,Kaplan:2009kr,Antipin:2009wr,Vecchi:2010jz,Kusafuka:2011fd,Sannino:2012wy}, as we shall see, leads to the expected slow running of the coupling constant towards the infrared also known as {\it walking} dynamics pioneered by Holdom  \cite{Holdom:1981rm,Holdom:1984sk} and also investigated in \cite{Yamawaki:1985zg,Appelquist:1986an,Appelquist:1988sr, Chivukula:1996kg}. The fixed points merge when the number of flavors reaches a critical value. Due to the calculable nature of the model we can provide a clear realization of the Miransky scaling \cite{Miransky:1984ef,Miransky:1988gk,Miransky:1996pd} and walking dynamics.  

The phenomenon of merging fixed points unveils itself here when going to the second order in perturbation theory. In fact, to
 the first order conformality can be lost, in our model, only because of the quantum corrections inducing spontaneous condensation of the scalar degrees of freedom \cite{Antipin:2011aa}.  Other quantities for walking dynamics such as the anomalous dimension of the fermion masses will also be determined precisely.

 In section \ref{Model} we review the model and the one-loop results. In section \ref{BOL} we setup the two-loop beta function analysis and discuss the fixed point structure. We also show that the IR stable fixed point merges with an UV fixed point as we decrease the number of flavors within the perturbative regime. We then move to section \ref{phys-merger} where we investigate in detail the physics of the merger. Here we determine precisely the conformal window as function of the number of flavors. Remarkably we are even able to determine the precise region of walking.  We show that the Miransky scaling emerges naturally in this model. We finally compute that anomalous dimension of the mass operator both for the UV and IR non-trivial fixed points. We briefly conclude in section \ref{conclu} while in the appendix \ref{full-betas} we provide the full beta function and anomalous dimension of the fermion mass operator of the theory at two loops and for any number of colors.

 \section{QCD with one Adjoint Fermion and Elementary Mesons}
\label{Model}
We want to investigate the dynamics of the following gauge-Yukawa theory: 
\ea{
\mathcal{L}&= 
\Tr \left[- \frac{1}{2} F^{\mu \nu}F_{\mu \nu} +i \bar{\lambda}  \slashed{D} \lambda+ \overline{Q} i \slashed{D} Q + \partial_\mu H ^\dagger \partial^\mu H + y_H \overline{Q} H Q  \right]- u_1 (\Tr [H ^\dagger H])^2 -u_2\Tr(H ^\dagger H )^2 \ . 
 \label{eq:Llsm}
}
The theory is, at the tree-level, conformally invariant.  With "$\Tr$" we indicate the trace operator over both color and flavor indices and ${D_\mu}$ is the covariant derivative respecting the gauge symmetries of the field on which it acts. The field content and symmetries are given in Table~\ref{FieldContent}, where the Dirac fermion is decomposed in its Weyl components $Q = (q, \widetilde{q}^*)$. The gauge coupling constant will be denoted by $g$. This theory is QCD with one adjoint Weyl fermion and elementary mesons. It has been chosen since, as we shall see, it allows for a very interesting perturbative gauge dynamics, and for its close resemblance to QCD. 
\begin{table}[b]
\vspace{-3mm}
\caption{Field content. The first three fields are Weyl spinors in the ($\frac{1}{2},0$) representation of the Lorentz group. $H$ is a complex scalar and $G_\mu$ are the gauge fields. $U(1)_{AF}$ is the extra anomaly free symmetry arising due to the presence of $\lambda$.}%
\vspace{-5mm}
\[ \begin{array}{c|c|c c c c} \hline \hline
{\rm Fields} &\left[ SU(N_c) \right] & SU(N_f)_L &SU(N_f)_R & U(1)_V& U(1)_{AF} \\ \hline 
\lambda & {\rm Adj} & 1 & 1 & 0 & 1 \\
 q &\yf &\overline{\yf }&1&~~\frac{N_f-N_c}{N_c} & - \frac{N_c}{N_f}\\
\widetilde{q}& \overline{\yf}&1 &  {\yf}& -\frac{N_f-N_c}{N_c}& - \frac{N_c}{N_f}\\
 \hline
  H & 1 & \yf & \overline{\yf} & 0 & \frac{2N_c}{N_f}\\
  G_\mu & \text{Adj} & 1 & 1 & 0 & 0 \\
   \hline \hline \end{array}%
\]%
\label{FieldContent}%
\vspace{-5mm}
\end{table}

Throughout this paper we will work with the rescaled couplings which allow for a finite Veneziano limit of the theory, i.e. $N_c, N_f \to \infty$ while keeping $x \equiv N_f/N_c$ fixed. The  appropriate rescaled couplings read
\eas{
a_g = \frac{g^2N_c}{(4 \pi)^2} \ ,~\ a_H = \frac{y_H^2N_c}{(4 \pi)^2}\ ,~ z_1 = \frac{u_1N_f^2}{(4 \pi)^2}\  , ~z_2 = \frac{u_2N_f}{(4 \pi)^2} \ .
}
If not otherwise stated, we will be working in the Veneziano limit.

This model was introduced in \cite{Antipin:2011aa}, where a one loop analysis was performed and showed to yield several interesting results. In this paper, we start the two-loop analysis of the theory where, as we will demonstrate, new important features emerge.
It is relevant for the completeness of this paper to summarize the most important features found at one loop, while referring the reader to \cite{Antipin:2011aa} for the details of the analysis (see also \cite{Grinstein:2011dq} for another model featuring similar properties).

\begin{itemize}
\item[i.] The theory has a perturbative, infrared stable fixed point (IRFP) in all couplings near the asymptotically free  boundary given by: 
\ea{\label{AF}
\overline{x} = 9/2 \ .
}
The perturbative nature of the fixed point (FP) is illustrated by showing the value of the couplings at the IRFP as a Taylor expansion around the asymptotically free boundary $\overline{x}$ to leading order: 
\eas{
a_g^* = \frac{22}{81}(\overline{x}-x), \ \ a_H^* = \frac{4}{27} (\overline{x}-x), \ \
z_1^* = \frac{-2\sqrt{19}+ \sqrt{2(8+3\sqrt{19})}}{27}(\overline{x}-x),  \ \ z_2^* = \frac{-1+\sqrt{19}}{27}(\overline{x}-x) \ .
}
We note that the existence of a perturbative stable IRFP for a generic non-supersymmetric gauge theory featuring scalars is not automatic \cite{Antipin:2011ny}. For instance, if we were to eliminate the adjoint Weyl fermion $\lambda$ and consider simply QCD with elementary mesons, such a perturbative fixed point would disappear.  In Fig.\ref{1LFP} we show the dependence of the  IRFP on $x$ to better appreciate its perturbative nature. 
\begin{figure}[b]
\centering
\includegraphics[width=0.6\columnwidth]{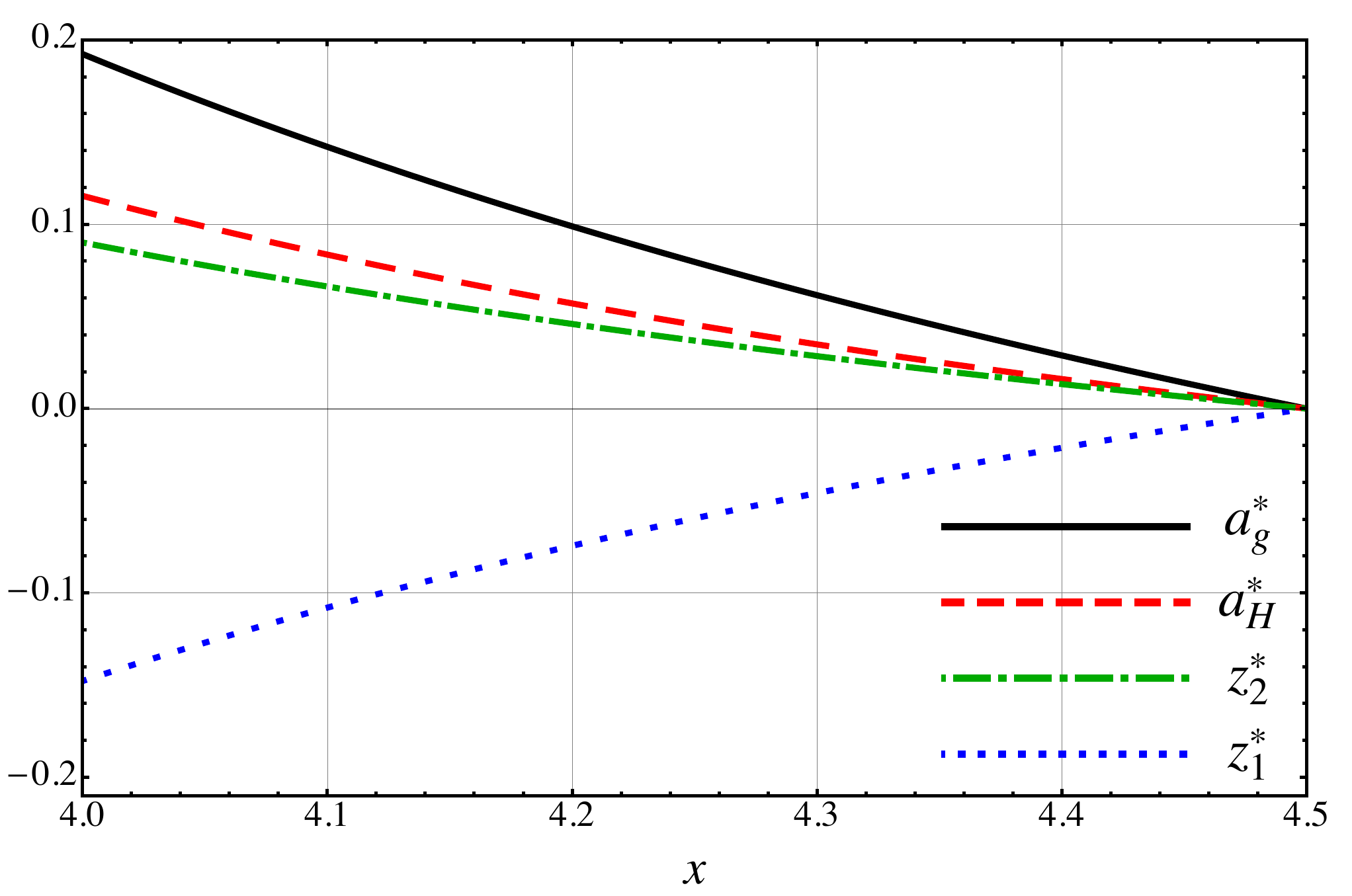}
 \caption{Fixed point solutions at one loop as a function of $x=N_f/N_c$.}
\label{1LFP}
\end{figure}

\item[ii.] Although there is a perturbatively stable IRFP not all the Renormalization Group (RG) trajectories will be able to end on it because of the presence of elementary mesons. In fact, the scalar potential can be unstable against a spontaneous chiral symmetry breaking phase transition induced by the Coleman-Weinberg (CW) phenomenon \cite{Coleman:1973jx}. This phenomenon is not encoded in the perturbative running of the couplings. A critical boundary (separatrix) was discovered in the parameter space of the theory separating the stable region from the unstable one. This is shown as a two-dimensional section of the parameter space in Fig.~\ref{RGFlow1}, where it is assumed that the gauge and Yukawa couplings have quasi-reached the FP.  
\begin{figure}[tb]
\centering
\includegraphics[width=0.5\columnwidth]{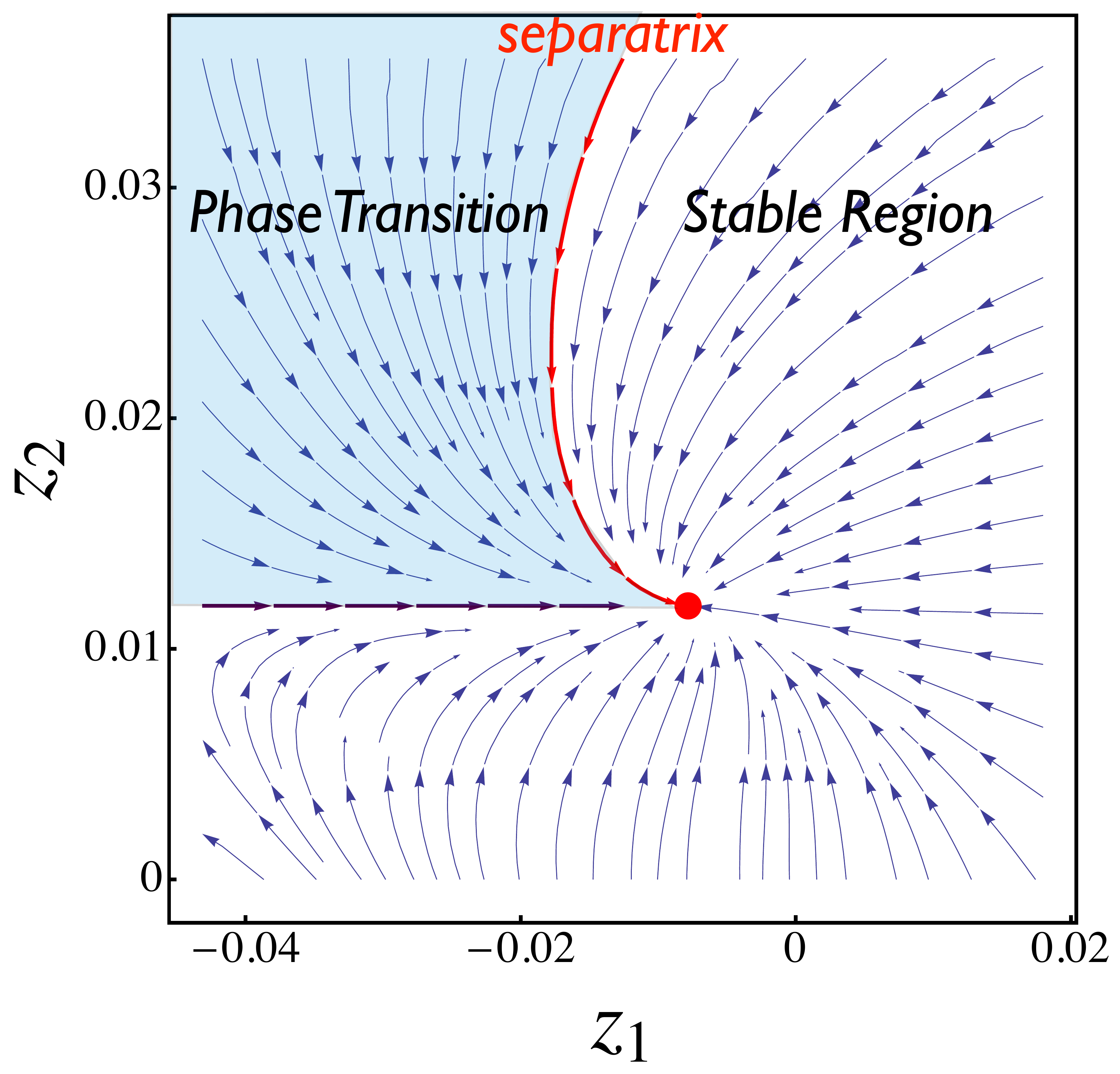}
 \caption{Showing the separation of phases of the RG flow, as discussed in the text,  for $x=4.4$. The other couplings, $a_g$ and $a_H$, are imagined to have quasi-reached the fixed point for}
\label{RGFlow1}
\end{figure}

\item[iii.]
In the stable region, i.e. where the CW does not lead to spontaneous symmetry breaking, the FP is reached and the theory exhibits large distance conformality. In the unstable region the effective potential develops a non-trivial minimum leading to chiral symmetry breaking to its maximal diagonal subgroup. The resulting perturbative spectrum consists of heavy fermions, heavy scalars, massless Nambu-Goldstone (NG) bosons and a pseudo-NG boson associated with spontaneous breaking of the dilatation symmetry.  {The gauge sector and $\lambda$-field are not affected by the symmetry breaking and emerge as a low-energy super Yang-Mills theory of which the condensate is computable nonperturbatively. The scale of the condensate/confinement of this remaining low energy theory can be shown to be exponentially lighter than any other generated mass scale of the theory. }

\item[iv.]
It has also been shown that the pseudo-NG boson, associated to dilatonic invariance, is the dilaton state of the theory at the quantum level. Its mass, to one loop order, was computed to be:
\ea{
m_d^2\Big |_{z_1=-z_2} = 4\mu_0^2 \left [4 z_2^2 - xa_H^2\right ]\exp \left ( -\frac{1}{2} - \frac{4z_2^2\ln(4z_2)-xa_H^2 \ln(x a_H)}{4 z_2^2 - xa_H^2} \right) \ ,
\label{md}
}
where $\mu_0$ is the scale at which the (bare) Lagrangian is defined and it is approximately the mass scale of the other heavy states. It was shown that $m_d^2$ can be made much lighter than the heavy mass scale by tuning $4 z_2^2 - xa_H^2 \ll a_H, z_2$, {since the heavy states acquire tree level masses proportional to $a_H$ and $z_2$. }
This tuning corresponds to RG trajectories closer and closer to the separatrix. A large scale separation between the different states of the theory is therefore obtained in the chirally  broken region by tuning the theory closer to the separatrix. 
\end{itemize}
We have thus constructed an explicit perturbative realization of conformal and near conformal dynamics. However, this realization differs from the one in QCD, since it is perturbative and furthermore the tuning needed to go in and out of the conformal window is in the couplings instead of the number of quark flavors. Furthermore, the primary condensation phenomenon involves a scalar field rather than the quark condensate. As we shall see in the following sections, this situation becomes more interesting when two loop corrections are included.

\section{The Two Loops Beta Functions and Emergence of New Fixed points}
\label{BOL}
It is interesting to go beyond one-loop since we know that higher order analyses often reveal several new important features \cite{Sannino:2010ca,DiChiara:2010xb,Mojaza:2010cm,Pica:2010mt,Pica:2010xq,Ryttov:2010iz,Sondergaard:2011ps,Ryttov:2012qu} not present at the lowest order. In fact, by carrying the analysis one step further in the perturbative expansion we will discover that the IRFP can also disappear, within the perturbative regime, by merging with a new ultraviolet fixed point as we decrease the number of quark flavors. The mechanism behind the loss of conformality is very different from the case in which this occurs via scalar condensation induced by quantum corrections. In this case, it is reasonable to expect the quarks to condense first and induce the scalar condensate. Of course, we have to work  in the region of the parameter space where the CW potential does not lead to a scalar condensate.  We have checked, by an explicit computation, that such a region of couplings exists.

We now therefore present the beta functions and their fixed point (FP) structure at two loops.
The two-loop beta functions have been derived using the formalism of \cite{Machacek:1983tz, Machacek:1983fi, Machacek:1984zw, Luo:2002ti}. In particular, the parametrization in \cite{Paterson:1980fc} was used to handle the scalar sector. The beta functions read:
\ea{
\label{betaag}\beta_{a_g} ={}& -\frac{2}{3} a_g^2 \left[11-2\ell-2 x+\left(34-16\ell-13 x\right)a_g +3x^2 a_H\right] \\[2mm]
\label{beta1H}\beta_{a_H} ={}& a_H  \left[ 2(x+1) a_H - 6 a_g +(8 x+5) a_g a_H+\frac{20 (x+\ell)-203}{6} a_g^2-8 x z_2 a_H-\frac{x(x+12)}{2} a_H^2+4 z_2^2\right] \\
\begin{split}
\label{beta1Y}\beta_{z_1} ={}& 4(z_1^2 +4  z_1 z_2+3 z_2^2+z_1 a_H) \\ 
&+2 \left(5 z_1 a_g a_H+2 x^2 a_H^3+x (4 z_2-3 z_1) a_H^2 -4 z_1^2 a_H-12 z_2^2 a_H-16 z_1 z_2 a_H-48 z_2^3-20 z_1 z_2^2\right) 
\end{split}\\[2mm]
\label{beta2Y} \beta_{z_2} ={}& 2 \left(2 z_2 a_H+4 z_2^2-x a_H^2\right)+2 \left(5 z_2 a_g a_H-2 x a_g a_H^2+2 x^2 a_H^3-3 x z_2 a_H^2-8 z_2^2 a_H-12 z_2^3\right)
}
The parameter $\ell$ is the number of $\lambda$ fields of the theory. In our case $\ell = 1$. We have kept $\ell$ as an external parameter, such that the results can be used for future research e.g. the model of QCD and mesons ($\ell =0$). 
Notice that the double trace coupling $z_1$ decouples from the beta functions of the remaining couplings  $a_g, a_H, z_2$ in the Veneziano limit. Moreover, $\beta(z_1)$ is a quadratic equation in $z_1$. 
Both of these properties actually hold to \emph{all orders} in perturbation theory in the Veneziano limit, as was shown in \cite{Pomoni:2008de}.
In Appendix \ref{full-betas} we provide the full $N_f$ dependent beta functions to two loops.
We have determined these expressions by implementing the above mentioned formalism in two independent ways for cross checking and have furthermore verified them by comparing with the partial two loop results obtained in \cite{Bardeen:1993pj} for a similar model without gauge interactions and the $\lambda$ field. We have also checked that the expressions are consistent with the Callan-Symanzik equation for the two loop effective potential \cite{Martin:2001vx}.

\begin{figure}[t]
\centering
\includegraphics[width=0.8\columnwidth]{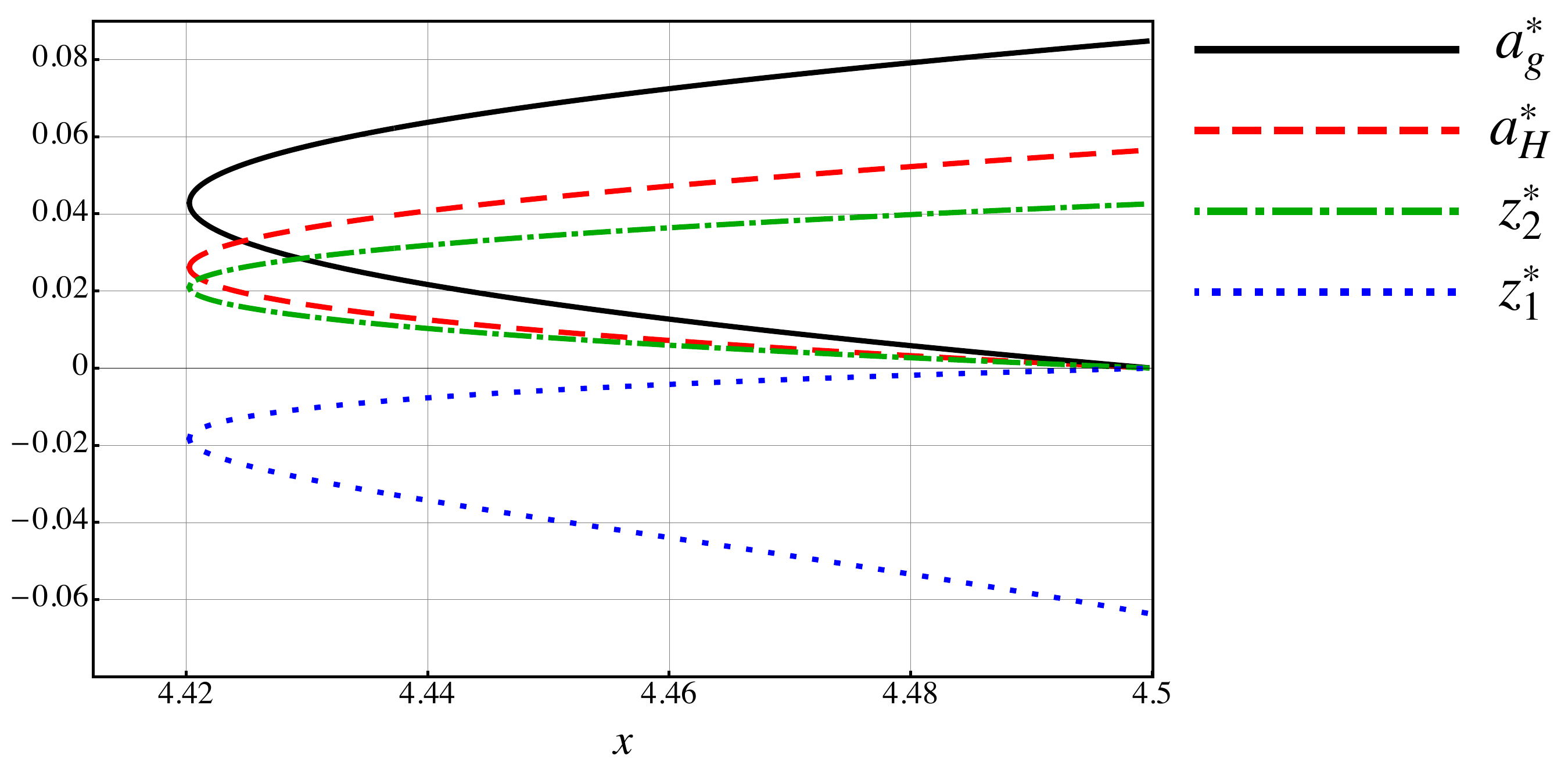} 
 \caption{Merging fixed points at two loops.}
\label{2loopFP}
\end{figure}

The system of beta functions has a rich fixed point structure which we will briefly discuss. 
Because of the decoupling and polynomial structure of $\beta(z_1)$ mentioned above, we know that it will always either have two real or no (i.e. two complex) FP solutions. For the remaining system of beta functions, we have found, depending on the value of $x$, at most six real solutions.  One of these corresponds to the perturbative IRFP discussed in Sec.\ref{Model}.  We focus our attention on the two loop fate of this IRFP as function of the number of flavors, i.e. $x$. To help the reader we provide its explicit two loop dependence on the number of flavors in Fig.\ref{2loopFP}. 
 It is clear from the figure that  the IRFP merges at a critical number of flavors  with a new ultraviolet (UV)FP appearing at two loops. Interestingly the UVFP appears to be perturbative in the same range of $x$ in which the IRFP exists.  Moreover, the merger occurs at $x^* = 4.42028$ which is quite close to the asymptotically free boundary for which the perturbative expansion is valid. It is therefore natural to expect that higher loop effects will not qualitatively affect this picture. This means that we discovered a conformal window as function of number of flavors accessible in perturbation theory. 

\section{Perturbative Miransky Scaling, Walking Dynamics and Critical Exponents}
\label{phys-merger}
Having investigated the fixed point structure of the theory at two loops we can turn our attention to the primary questions we address here. What kind of near-conformal dynamics appears as we change the number of flavors?  Does the theory jump or walk?  We will demonstrate that: a) The investigated gauge theory displays a smooth phase transition as a function of the number of flavors when crossing the lower boundary of the conformal window; b) There is a well defined region of walking dynamics; c) We can reliably compute the relevant critical exponents of the theory such as the anomalous dimension of the fermion mass operator in the entire conformal window.

We have already discussed that the IRFP as function of the number of flavors merges with the UV one, and argued that this happened within the perturbative regime of the theory. It is therefore timely to start a careful perturbative investigation of the physical properties of the theory around the lower boundary of the conformal window delimited by  the merging of the IRFP and UVFP occurring at $x^* = 4.42028$ corresponding to a critical number of flavors $N_f^*=x^* N_c $. Asymptotic freedom is lost for $\bar{N}_f=\bar{x}N_c = 9N_c/2$. We will also consider what happens when we lower $x$ past the point of the merger in order to demonstrate the emergence of walking dynamics.

\begin{figure}[bt]
\centering
\includegraphics[width=0.55\columnwidth]{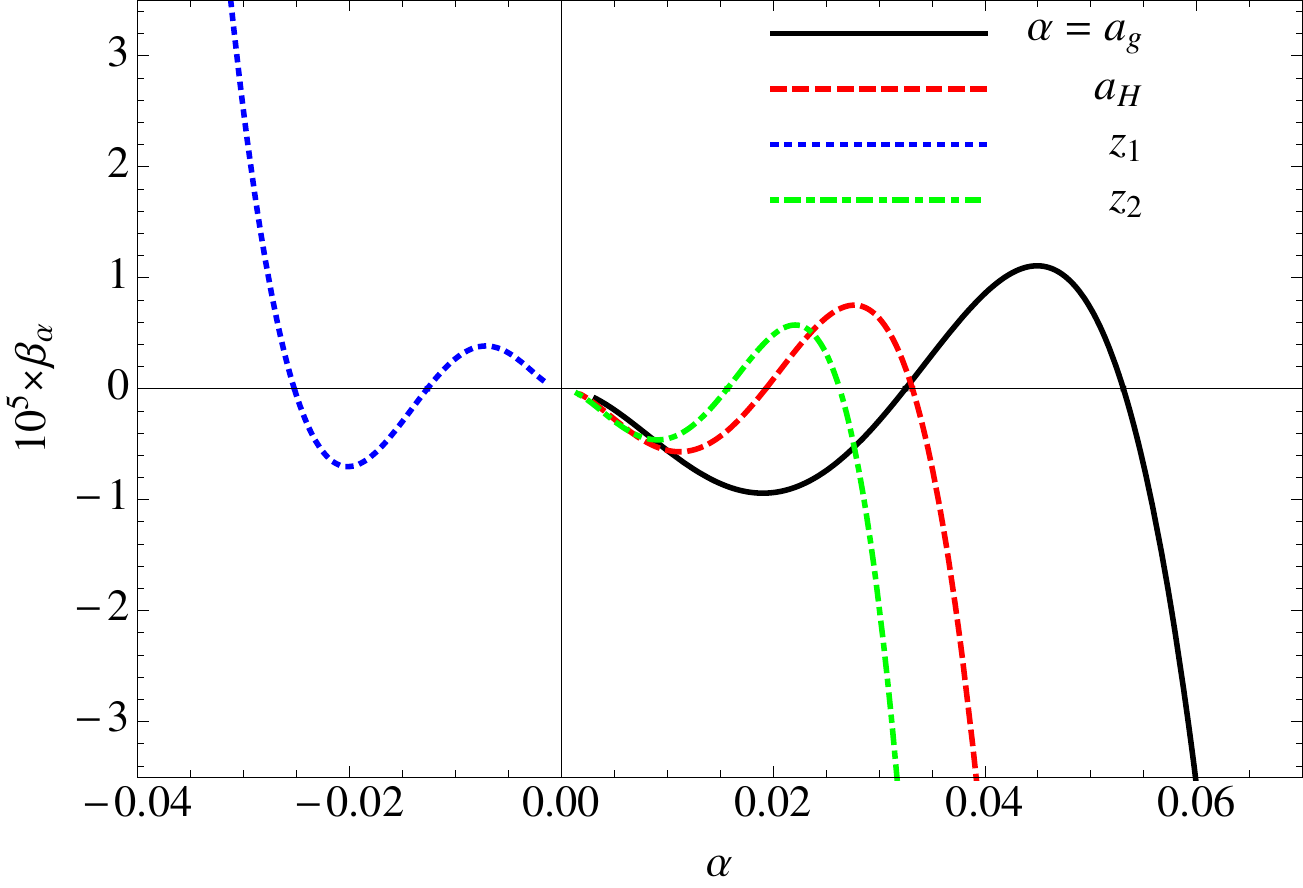} 
 \caption{Beta functions $\beta_\alpha(\alpha)$ for $\alpha=a_g,a_H,z_1,z_2$ and for $x=4.425$. Each beta function features two nontrivial FPs, one in the UV, which is unstable under small changes in the values of the four couplings, and one stable in the IR.}
\label{fig:betay}
\end{figure}

Within the conformal window, i.e. $x^*<x<\bar{x}$ the physics is controlled by the nontrivial IRFP point when starting the flow in the ultraviolet near the trivial UVFP. In Fig.~\ref{fig:betay} we show the different $\beta_\alpha(\alpha)$ for $\alpha=a_g,a_H,z_1,z_2$ at $x=4.425$ obtained by solving the RG equations numerically\footnote{Specifically each of the four betas is composed of three parametric plots of $\beta_\alpha(\mu),\alpha(\mu)$, determined by choosing different initial conditions, respectively one close to the origin, and two close to the UV FP but on different branches of $\beta_\alpha$, and for a large range of the renormalization scale $\mu$. 
}.

\begin{figure}[bt]
\centering
\includegraphics[width=0.55\columnwidth]{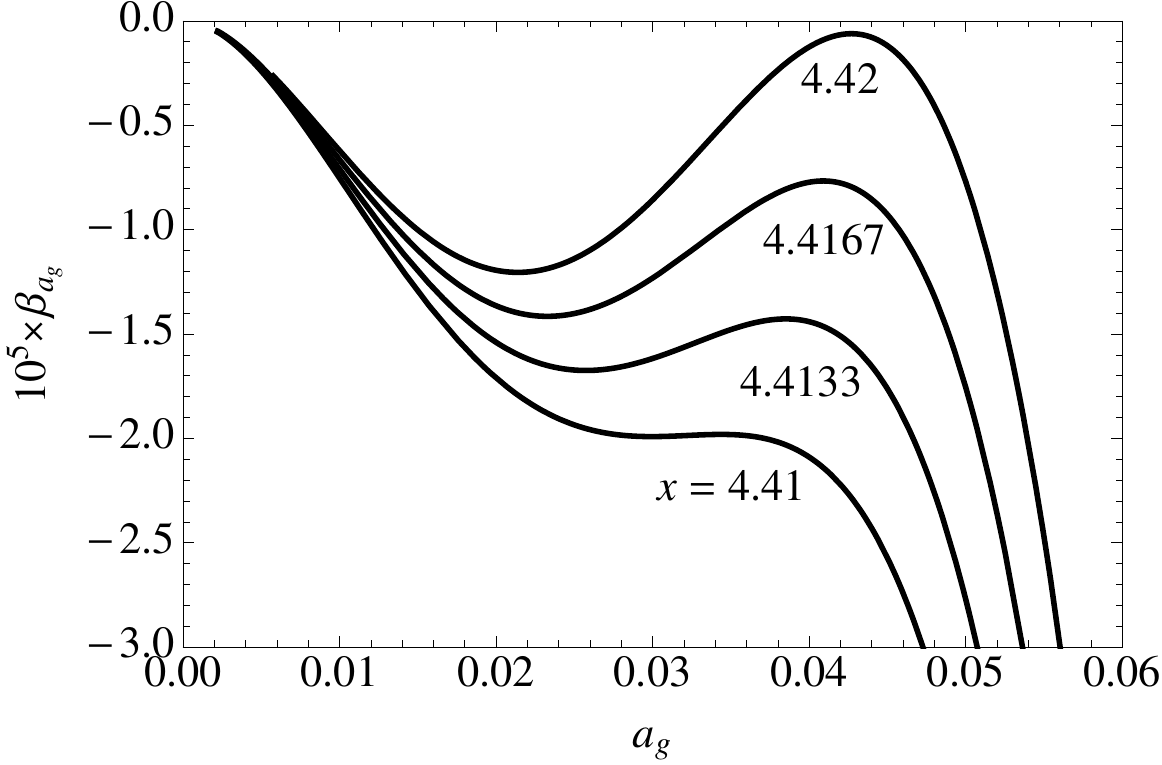} 
 \caption{The beta function $\beta_{a_g}(a_g)$  plotted for $x=4.42,4.4167,4.4133,4.41$. The local maximum of  $\beta_{a_g}$ disappears for $x\simeq 4.41$.}
\label{fig:betaw}
\end{figure}

When $x$ is only slightly lower than $x^*$,  we find that this situation is mostly unchanged, since, as we shall demonstrate, the dynamics is controlled by the quasi FPs. The beta function $\beta_{a_g}(a_g)$ is shown in Fig.~\ref{fig:betaw} for different values of $x$ near the merger.

Below the merger the theory must develop a physical scale $\Lambda$  and we introduce therefore the {\it renormalization group time} $t \equiv \log \frac{\Lambda}{\mu_0}$ it takes for the theory to flow from a UV scale $\mu_0$ to the infrared scale $\Lambda$. 

\begin{figure}[b]
\centering
\includegraphics[width=0.55\columnwidth]{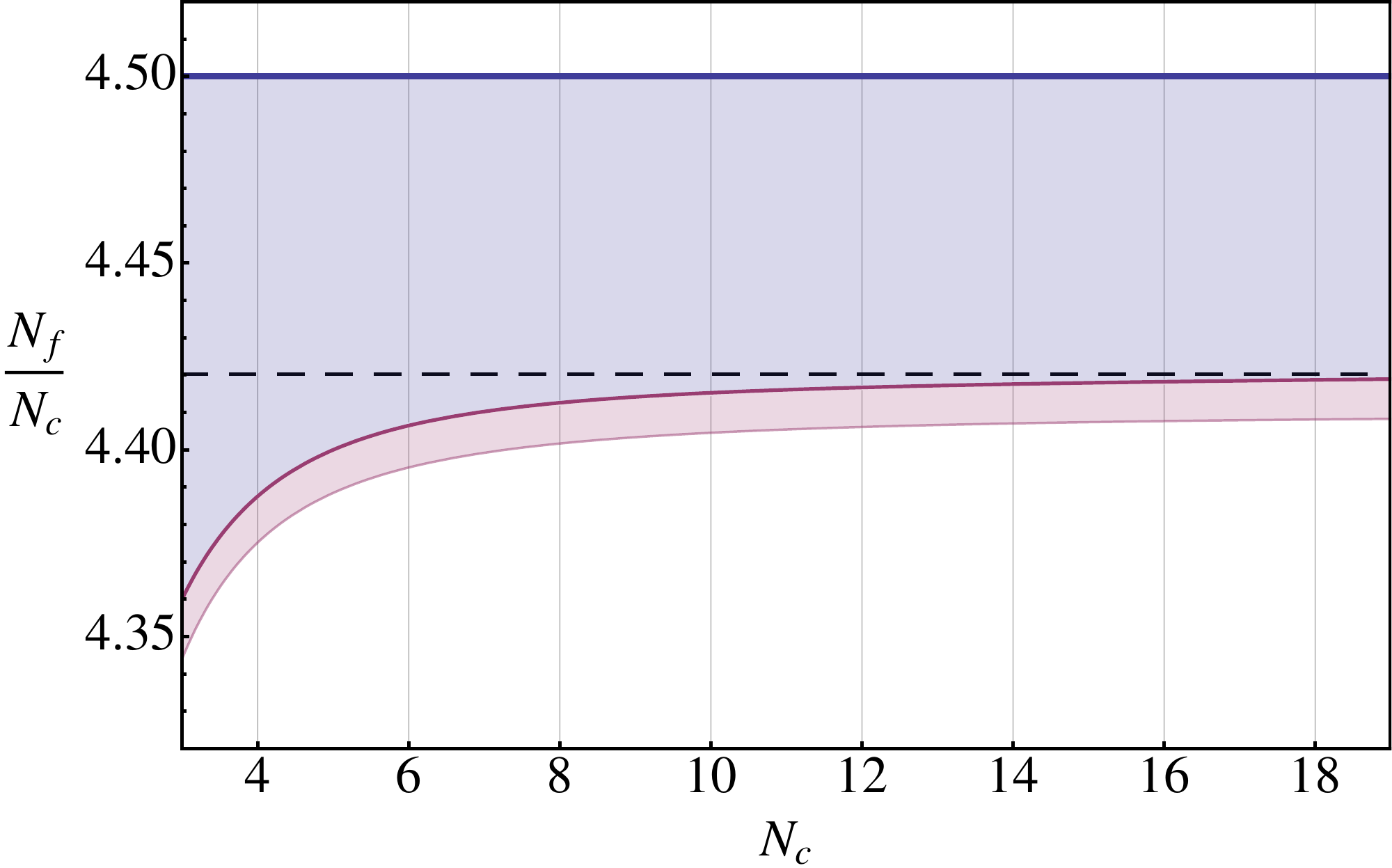}
 \caption{Phase diagram including the conformal window (light blue shaded region) and walking region (pink shaded region) of the theory as function of the number of flavors and colors. The dashed line corresponds to the 
 large $N_c$ limit of the lower boundary of the conformal window.}
\label{Conformal_Window}
\end{figure}

It is clear from Fig.~\ref{fig:betaw} that a local maximum of  $\beta_{a_g}$ is present at a value of the coupling close to the merger FP value. The range of values of $x<x^*$ for which $\beta_{a_g}$ features a local maximum away from the trivial FP  defines concretely the walking dynamics regime. Below the value of $x$ where the local maximum disappears  the theory no longer walks but runs. From Fig.~\ref{fig:betaw} it is evident that this happens for $x \approx 4.41$. We have thus managed to estimate the \emph{walking window} of the theory:
\begin{alignat*}{2}
0 &< x \leq 4.41 &\quad & \text{Running Dynamics} \\
4.41 &< x \leq 4.42 & & \text{Walking Dynamics}\\
4.42 &< x < 4.5 & & \text{Conformal Window}\\
4.5 &\leq x  & & \text{IR free}
\end{alignat*}
The phase diagram including the conformal window (light blue shaded region) and walking region (pink shaded region) as a function of the number of flavors and colors is reported in Fig.~\ref{Conformal_Window}, taking into account the full dependence on the number of colors  $N_c$. Deviations from the large $N_c$ limit (dashed line) are sizable at small  $N_c$ as it can be seen from the figure.  

\begin{figure}[bt]
\centering
\includegraphics[width=0.55\columnwidth]{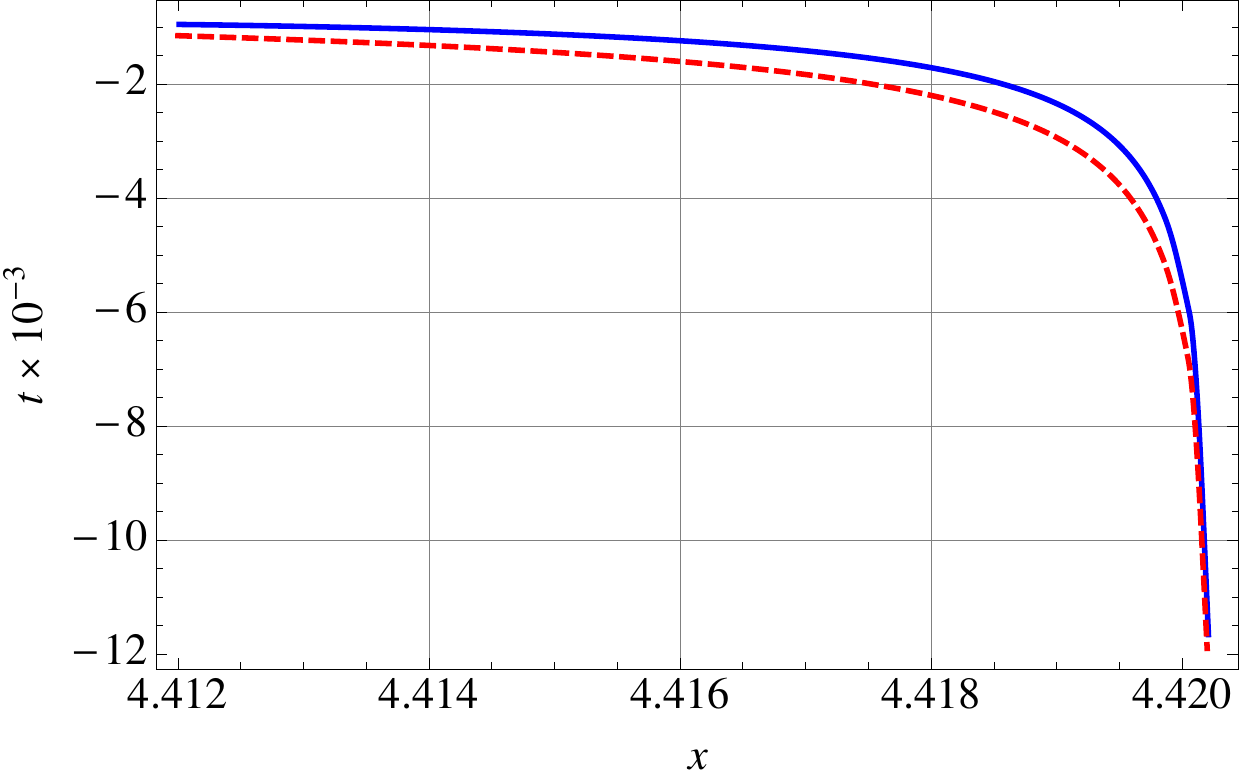} 
 \caption{The Miransky scaling function (red dashed) in Eq.~\ref{Msc} plotted together with the numerically precise evaluation of the integral in Eq.~\ref{tbg} (solid blue) for the interval  $4.41201\leq x\leq4.42021$. The Miransky scaling approximation becomes increasingly accurate as $x$ gets closer to $x^*$.}
\label{fig:Msc}
\end{figure}

We can then determine $\Lambda$ in the walking regime as follows\ea{\label{tbg}
\log \frac{\Lambda}{\mu_0}= \int_{\mu_0}^{\Lambda}\frac{\id \mu}{\mu} =\ \int_{a_g(\mu_0)}^{a_g(\Lambda)}\frac{\id a_g}{\beta_{a_g}(a_g,k_1,k_2,k_3)}\ .
}
where $k_i$ with $i=1,2,3$ are the integration constants for the other couplings solved as a function of $a_g$. 
Within the walking region the theory will spend most of the RG time near the quasi-FP $a_g^*$, corresponding to a local maximum of $\beta_{a_g}$. It is therefore natural to pick $\mu_0$ such that $a_g(\mu_0)=a_g^*$ \cite{Braun:2010qs}. In the deep infrared the $a_g$ coupling diverges defining $\Lambda$. We have  computed the last integral in Eq.~\ref{tbg} numerically, which is shown in Fig.~\ref{fig:Msc}.  It is interesting to compare the numerical result with the leading order analytic contribution of $\beta_{a_g}$ to the integral in Eq.~\ref{tbg}, which is determined by expanding $\beta_{a_g}$ around its local maximum:
\ea{
\int_{a_g^*}^{\infty } \frac{1}{\beta _0+\frac{\beta _2
   \left(a_g-a_g^*\right){}^2}{2!}} \, da_g=-\frac{\pi }{\sqrt{\beta _0 \beta _2}},
\label{parint}   
}
with
\ea{
\left.\partial_{a_g} \beta _{a_g}\right|_{a_g^*}=0,\quad \beta_0\equiv\beta_{a_g}(a_g^*),\quad \beta_2\equiv\left.\partial_{a_g,a_g} \beta _{a_g}\right|_{a_g^*} .
}
The quantities $\beta_{0,2}$ depend on $x$ and, of course, also on the $k_i$ integration constants. We evaluated them for $4.41201\leq x\leq4.42021$ in increments in $x$ of $10^{-4}$ and define $\beta_{0,2}(x)$ as the best fit fourth order polynomials in $x-x^*$ to the respective data points \footnote{Higher order polynomials do not change our results.}. The resulting leading order contribution in $x$ we obtain for the RG time is
\ea{
\log \frac{\Lambda}{\mu_0} \simeq-\frac{\pi}{c_1 \sqrt{x^*-x}}\ ,\quad c_1=2.99 \times 10^{-2}\ .
\label{Msc}   
}
The result above shows that the model realizes   the Miransky scaling   \cite{Miransky:1988gk}. The Miransky scaling function in Eq.~\ref{Msc} is plotted in Fig.~\ref{fig:Msc} together with the precise numerical evaluation of the integral in Eq.~\ref{tbg} for the interval  $4.41201\leq x\leq4.42021$, which covers most of the walking region in $x$.
It is clear from  Fig.~\ref{fig:Msc} that the Miransky scaling approximation becomes increasingly accurate as $x$ gets closer to $x^*$. This is an expected result, since for $x\rightarrow \left(x^*\right)^-$ higher order corrections to the inverse parabolic approximation of $\beta_{a_g}^{-1}$ become negligible.

\begin{figure}[h]
\centering
\includegraphics[width=0.49\columnwidth]{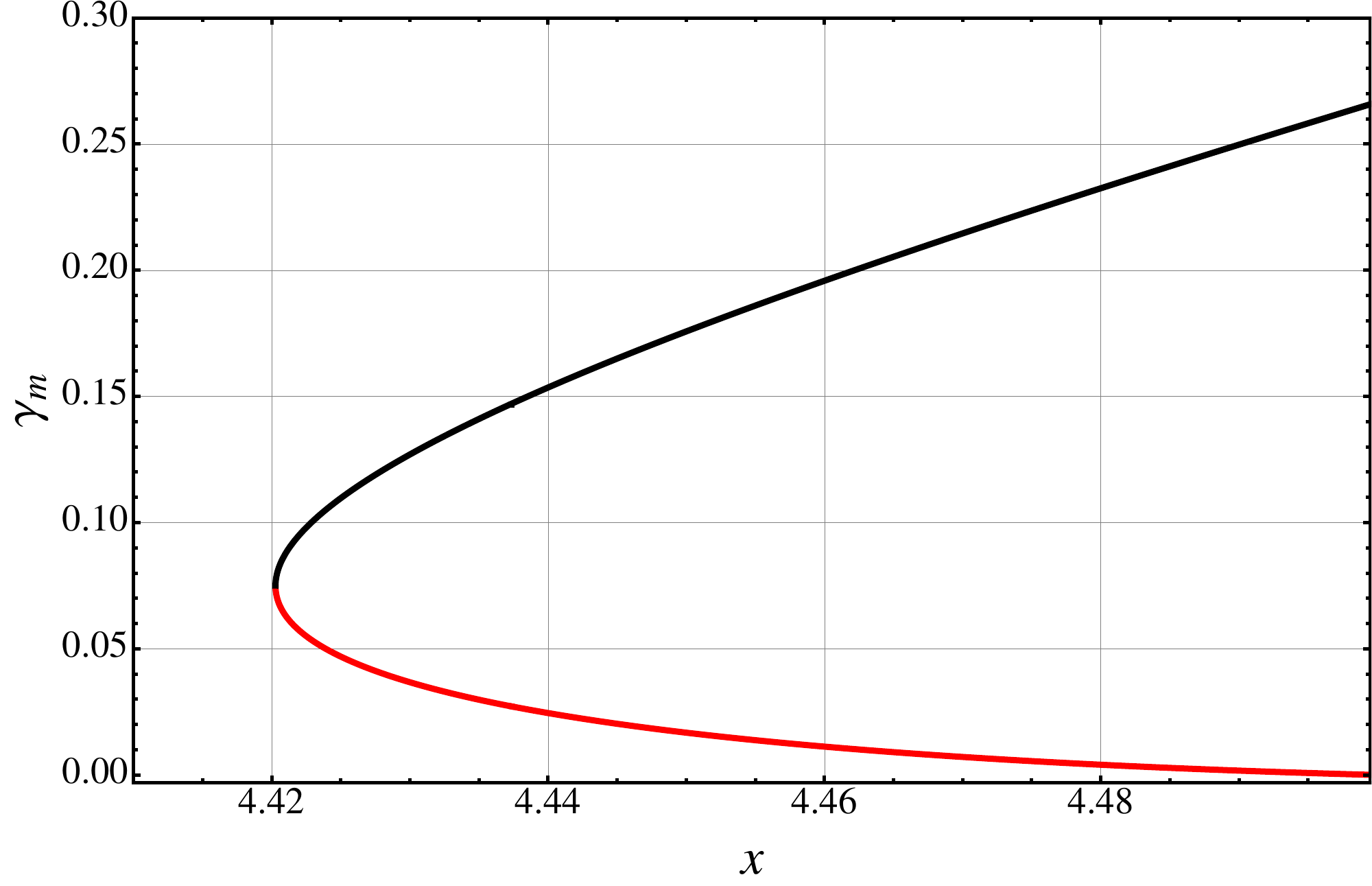} 
 \caption{ Anomalous dimension of the fermion mass at the UVFP (upper black) and IRFP (lower red). }
\label{fig:gamma}
\end{figure}

 So far we have shown that the model possesses a smooth conformal phase transition when crossing the lower boundary of the conformal window as function of the number of flavors. At the FPs anomalous dimensions are expected to be physical quantities identifiable with the critical exponents of the theory. These are important quantities governing the scaling properties of relevant correlators and therefore deserve to be investigated. A particularly interesting one is the anomalous dimension of the fermion mass operator. This quantity  evaluated at the merger is very relevant for walking dynamics 
\ea{
\gamma_m\equiv -\frac{d\log m}{d\log\mu}\hspace{1mm}, \qquad \Delta_{\bar{\psi}\psi}=3-\gamma_m \ .
}
The direct contributions to $\gamma_m$ at the one and two-loop order are \cite{Luo:2002ti}:
\ea{
\label{g1}\gamma_m^{(1)}&=3 a_g-x a_H\ ,  \\ 
\label{g2}\gamma_m^{(2)}&=\frac{1}{12} \left(96 x a_g a_H+(203-20 x-20\ell) a_g^2+3 x (x+6) a_H^2\right) \ .
}
As in the expressions for the beta functions $\ell$ denotes the number of $\lambda$ fields.
For completeness, the full two-loop expression for the anomalous dimension, not truncated to the Veneziano limit, is provided in appendix \ref{full-betas}. 
Notice that the quartic interactions do not contribute to the anomalous dimension at two-loop order.  In Fig.~\ref{fig:gamma} the red and black lines corresponds to $\gamma_m= \gamma_m^{(1)}+\gamma_m^{(2)}$ at the IRFP (red) and UVFP (black) given in Fig.~\ref{2loopFP}.  The physical dimension of the fermion mass operator at the merger reads $\Delta_{\bar{\psi}\psi}\approx 2.92$. In the walking regime  the physics is dominated by the nearby FPs and therefore the relevant anomalous dimension  is $\gamma_{\rm merger}\approx 0.08$.  

Remarkably this model features all the salient properties of walking dynamics. Of course, larger values of the anomalous dimensions can arise for theories which are strongly coupled at the phase transition point. Nevertheless it is interesting to have shown that walking dynamics is actually realized in asymptotically free four-dimensional gauge theories.

\section{Conclusions}
\label{conclu}
  We introduced a relevant four dimensional gauge theory template which has allowed us to determine its conformal and walking windows as functions of the number of flavors within perturbation theory. We have shown that conformality is lost due to the merging of an IR and UV fixed points leading to Miransky scaling. We were able to define the walking regime of the theory and compute the dependence on the number of flavors of the anomalous dimension of the fermion mass. 

Walking dynamics is required as an essential feature for extensions of the standard model featuring new asymptotically free gauge theories. However,  many of its required properties are either conjectured or investigated using model computations. Here we have first introduced a relevant asymptotically free gauge theory featuring $N_f$ flavors in the fundamental representation of an $SU(N_c)$ gauge group, like QCD, and then added a Weyl fermion in the adjoint together with meson-like fundamental scalars.  

At the two-loop level, and in the Veneziano limit, we discovered the existence of several fixed points of which one is all directions stable in the infrared. We showed that this fixed point is lost, within the perturbatively trustable regime, by merging with another ultraviolet fixed point when varying the number of quark flavors.  We naturally recover the  Miransky scaling and determine its properties. The theory is manifestly of walking type and the theory allows us to determine the associated walking window in the number of colors and flavors. We find that the walking window is about 12\% of the conformal window at large $N$.  Furthermore, we determined also other interesting quantities for walking dynamics such as the anomalous dimension of the fermion mass.

\appendix

\section{The full two-loop $\beta$-functions and anomalous dimension}\label{full-betas}
For completeness, we provide the $\beta$-functions of the theory and the anomalous dimension of the fermion mass to the full two loop order. The parameter $\ell$ denotes the number of $\lambda$ fields.

\ea{
\beta_{a_g}={}&\frac{2}{3}a_g^2(2x-11+2\ell) + a_g^3\left(\frac{2}{3}(13x-34+16\ell)-\frac{2x^3}{N_f^2}\right)-2x^2a_g^2a_H\\
\begin{split}
\beta_{a_H}={}&-6a_ga_H\left(1-\frac{x^2}{N_f^2}\right)+2a_H^2(x+1)+a_g^2a_H\left(\frac{1}{6}(20x-203+20\ell)-\frac{2x^2(5x-53+5\ell)}{3N_f^2}-\frac{3x^4}{2N_f^4}\right)\\
&+a_ga_H^2\left(8x+5-\frac{x^2(8x+5)}{N_f^2}\right) - a_H^3\left(\frac{x}{2}(x+12)-\frac{4x^2}{N_f^2}\right) -8xa_H^2\left(\frac{2}{N_f^2}z_1+\left(1+\frac{1}{N_f^2}\right)z_2\right)\\
&+4a_H\left(\left(\frac{1}{N_f^2}+\frac{1}{N_f^4}\right)z_1^2+\frac{4}{N_f^2}z_1z_2+\left(1+\frac{1}{N_f^2}\right)z_2^2\right)
\end{split}\\
\begin{split}
\beta_{z_1}={}&4a_Hz_1+4z_1^2\left(1+\frac{4}{N_f^2}\right) + 16z_1z_2 + 12z_2^2+10a_ga_Hz_1\left(1-\frac{x^2}{N_f^2}\right)+4x^2a_H^3-2xa_H^2(3z_1-4z_2)\\
&-8a_H\left(\left(1+\frac{4}{n_f^2}\right)z_1^2+4z_1z_2+3z_2^2\right)-24z_1^3\left(\frac{3}{N_f^2}+\frac{7}{N_f^4}\right)-\frac{352}{N_f^2}z_1^2z_2-8z_1z_2^2\left(5-\frac{41}{N_f^2}\right)-96z_2^3
\end{split}\\
\begin{split}
\beta_{z_2}={}&-2xa_H^2+4a_Hz_2+\frac{24}{N_f^2}z_1z_2+8z_2^2-4xa_ga_H^2\left(1-\frac{x^2}{N_f^2}\right)+10a_ga_Hz_2\left(1-\frac{x^2}{N_f^2}\right)+4x^2a_H^3\\
&+2xa_H^2\left(\frac{4}{N_f^2}z_1-3z_2\right)-16a_H\left(\frac{3}{N_f^2}z_1z_2+z_2^2\right)-8z_1^2z_2\left(\frac{5}{N_f^2}+\frac{41}{N_f^4}\right)-\frac{352}{N_f^2}z_1z_2^2-24z_2^3\left(1+\frac{5}{N_f^2}\right)
\end{split}\\
\begin{split}
\gamma_m ={}&3 a_g \left(1-\frac{x^2}{N_f^2}\right)-x a_H+a_g^2 \left(\frac{1}{12} (203-20 x-20 \ell )-\frac{x^2 (53-5 x-5 \ell )}{3 N_f^2}+\frac{3x^4}{4 N_f^4}\right)\\
&+8 x a_g a_H \left(1-\frac{x^2}{N_f^2}\right)+\frac{1}{4} a_H^2 x (6+x)
\end{split}
}


\begin{thebibliography}{99}

\bibitem{Sannino:2009za}
 F.~Sannino,
 Acta Phys.\ Polon.\  B {\bf 40}, 3533 (2009)
 [arXiv:0911.0931 [hep-ph]].

\bibitem{Channuie:2011rq}
 P.~Channuie, J.~J.~Joergensen and F.~Sannino,
 JCAP {\bf 1105}, 007 (2011)
 [arXiv:1102.2898 [hep-ph]].

\bibitem{Bezrukov:2011mv}
 F.~Bezrukov, P.~Channuie, J.~J.~Joergensen and F.~Sannino,
 arXiv:1112.4054 [hep-ph].

\bibitem{Hooft:2010nc}
 G.~'.~Hooft,
 arXiv:1011.0061 [gr-qc].

\bibitem{Holdom:1981rm}
 B.~Holdom,
 Phys.\ Rev.\  D {\bf 24}, 1441 (1981).

\bibitem{Holdom:1984sk}
 B.~Holdom,
 Phys.\ Lett.\  B {\bf 150}, 301 (1985).

\bibitem{Yamawaki:1985zg}
 K.~Yamawaki, M.~Bando and K.~i.~Matumoto,
 Phys.\ Rev.\ Lett.\  {\bf 56}, 1335 (1986).

\bibitem{Appelquist:1986an}
 T.~W.~Appelquist, D.~Karabali and L.~C.~R.~Wijewardhana,
 Phys.\ Rev.\ Lett.\  {\bf 57}, 957 (1986).

\bibitem{Sannino:2012wy}
 F.~Sannino,
 arXiv:1205.4246 [hep-ph].

\bibitem{Miransky:1984ef}
 V.~A.~Miransky,
 Nuovo Cim.\  A {\bf 90}, 149 (1985).

\bibitem{Miransky:1988gk}
 V.~A.~Miransky and K.~Yamawaki,
 Mod.\ Phys.\ Lett.\  A {\bf 4}, 129 (1989).

\bibitem{Miransky:1996pd}
 V.~A.~Miransky and K.~Yamawaki,
 Phys.\ Rev.\  D {\bf 55}, 5051 (1997)
 [Erratum-ibid.\  D {\bf 56}, 3768 (1997)]
 [arXiv:hep-th/9611142].

\bibitem{deForcrand:2012se} 
  P.~de Forcrand, M.~Pepe and U.~-J.~Wiese,
  arXiv:1204.4913 [hep-lat].

\bibitem{Grinstein:2011dq}
 B.~Grinstein and P.~Uttayarat,
 JHEP {\bf 1107}, 038 (2011)
 [arXiv:1105.2370 [hep-ph]].

\bibitem{Antipin:2011aa}
 O.~Antipin, M.~Mojaza and F.~Sannino,
 Phys.\ Lett.\  B {\bf 712}, 119 (2012)
 [arXiv:1107.2932 [hep-ph]].

\bibitem{Mojaza:2011rw} 
  M.~Mojaza, M.~Nardecchia, C.~Pica and F.~Sannino,
  Phys.\ Rev.\ D {\bf 83}, 065022 (2011)
  [arXiv:1101.1522 [hep-th]].
  
\bibitem{Coleman:1973jx}
 S.~R.~Coleman and E.~J.~Weinberg,
 Phys.\ Rev.\  D {\bf 7}, 1888 (1973).

\bibitem{Gies:2005as}
 H.~Gies and J.~Jaeckel,
 Eur.\ Phys.\ J.\  C {\bf 46}, 433 (2006)
 [arXiv:hep-ph/0507171].

\bibitem{Kaplan:2009kr}
 D.~B.~Kaplan, J.~W.~Lee, D.~T.~Son and M.~A.~Stephanov,
 Phys.\ Rev.\  D {\bf 80}, 125005 (2009)
 [arXiv:0905.4752 [hep-th]].

\bibitem{Antipin:2009wr}
 O.~Antipin and K.~Tuominen,
 Phys.\ Rev.\  D {\bf 81}, 076011 (2010)
 [arXiv:0909.4879 [hep-ph]].

\bibitem{Vecchi:2010jz}
 L.~Vecchi,
 Phys.\ Rev.\  D {\bf 82}, 045013 (2010)
 [arXiv:1004.2063 [hep-th]].

\bibitem{Kusafuka:2011fd}
 Y.~Kusafuka and H.~Terao,
 Phys.\ Rev.\  D {\bf 84}, 125006 (2011)
 [arXiv:1104.3606 [hep-ph]].

\bibitem{Appelquist:1988sr}
 T.~Appelquist, D.~Nash and L.~C.~R.~Wijewardhana,
 Phys.\ Rev.\ Lett.\  {\bf 60}, 2575 (1988).

\bibitem{Chivukula:1996kg}
 R.~S.~Chivukula,
 Phys.\ Rev.\  D {\bf 55}, 5238 (1997)
 [arXiv:hep-ph/9612267].

\bibitem{Antipin:2011ny}
 O.~Antipin, M.~Mojaza, C.~Pica and F.~Sannino,
 arXiv:1105.1510 [hep-th].

\bibitem{Sannino:2010ca}
 F.~Sannino,
 Phys.\ Rev.\  D {\bf 82}, 081701 (2010)
 [arXiv:1006.0207 [hep-lat]].

\bibitem{DiChiara:2010xb}
 S.~Di Chiara, C.~Pica and F.~Sannino,
 Phys.\ Lett.\  B {\bf 700}, 229 (2011)
 [arXiv:1008.1267 [hep-ph]].

\bibitem{Mojaza:2010cm}
 M.~Mojaza, C.~Pica and F.~Sannino,
 Phys.\ Rev.\  D {\bf 82}, 116009 (2010)
 [arXiv:1010.4798 [hep-ph]].

\bibitem{Pica:2010mt}
 C.~Pica and F.~Sannino,
 Phys.\ Rev.\  D {\bf 83}, 116001 (2011)
 [arXiv:1011.3832 [hep-ph]].

\bibitem{Pica:2010xq}
 C.~Pica and F.~Sannino,
 Phys.\ Rev.\  D {\bf 83}, 035013 (2011)
 [arXiv:1011.5917 [hep-ph]].

\bibitem{Ryttov:2010iz}
 T.~A.~Ryttov and R.~Shrock,
 Phys.\ Rev.\  D {\bf 83}, 056011 (2011)
 [arXiv:1011.4542 [hep-ph]].

\bibitem{Sondergaard:2011ps}
 U.~I.~Sondergaard, C.~Pica and F.~Sannino,
 Phys.\ Rev.\  D {\bf 84}, 075022 (2011)
 [arXiv:1107.1802 [hep-ph]].

\bibitem{Ryttov:2012qu}
 T.~A.~Ryttov and R.~Shrock,
 Phys.\ Rev.\  D {\bf 85}, 076009 (2012)
 [arXiv:1202.1297 [hep-ph]].

\bibitem{Machacek:1983tz}
 M.~E.~Machacek and M.~T.~Vaughn,
 Nucl.\ Phys.\  B {\bf 222}, 83 (1983).

\bibitem{Machacek:1983fi}
 M.~E.~Machacek and M.~T.~Vaughn,
 Nucl.\ Phys.\  B {\bf 236}, 221 (1984).

\bibitem{Machacek:1984zw}
 M.~E.~Machacek and M.~T.~Vaughn,
 Nucl.\ Phys.\  B {\bf 249}, 70 (1985).

\bibitem{Luo:2002ti}
 M.~x.~Luo, H.~w.~Wang and Y.~Xiao,
 Phys.\ Rev.\  D {\bf 67}, 065019 (2003)
 [arXiv:hep-ph/0211440].

\bibitem{Paterson:1980fc}
 A.~J.~Paterson,
 Nucl.\ Phys.\  B {\bf 190}, 188 (1981).

\bibitem{Pomoni:2008de}
 E.~Pomoni and L.~Rastelli,
 JHEP {\bf 0904}, 020 (2009)
 [arXiv:0805.2261 [hep-th]].

\bibitem{Bardeen:1993pj}
 W.~A.~Bardeen, C.~T.~Hill and D.~U.~Jungnickel,
 Phys.\ Rev.\  D {\bf 49}, 1437 (1994)
 [arXiv:hep-th/9307193].

\bibitem{Martin:2001vx}
 S.~P.~Martin,
 Phys.\ Rev.\  D {\bf 65}, 116003 (2002)
 [arXiv:hep-ph/0111209].

\bibitem{Braun:2010qs}
 J.~Braun, C.~S.~Fischer and H.~Gies,
 Phys.\ Rev.\  D {\bf 84}, 034045 (2011)
 [arXiv:1012.4279 [hep-ph]].

\end{thebibliography}
\end{document}